\documentclass[twocolumn,aps,prl,groupedaddress]{revtex4}
\usepackage{amssymb}
\usepackage{amsmath}
\usepackage{color}
\usepackage{graphicx}
\usepackage{marvosym}

\newcommand{\be}{\begin{equation}}
\newcommand{\ee}{\end{equation}}
\newcommand{\nn}{\nonumber\\}

\newcommand{\la}{\langle}
\newcommand{\ra}{\rangle}
\newcommand{\lb}{\left[}
\newcommand{\rb}{\right]}
\newcommand{\lp}{\left(}
\newcommand{\rp}{\right)}

\newcommand{\sgn}{{\rm sgn}\,}

\renewcommand{\Re}{{\rm Re}\,}
\renewcommand{\vec}[1]{{\bf #1}}

\begin{document}

\title{Plasmon geometric phase and plasmon Hall shift}
\author{Li-kun Shi$^{1}$ and Justin C. W. Song$^{1,2}$}
\email{justinsong@ntu.edu.sg}
\affiliation{$^1$Institute of High Performance Computing, Agency for Science, Technology, \& Research, Singapore 138632}
\affiliation{$^2$Division of Physics and Applied Physics, Nanyang Technological University, Singapore 637371}

\begin{abstract}
The collective plasmonic modes of a metal comprise a pattern of charge density and tightly-bound electric fields that oscillate in lock-step to yield enhanced light-matter interaction. 
Here we show that metals with non-zero Hall conductivity host plasmons with a fine internal structure: they are characterized by a current density configuration that sharply departs from that of ordinary zero Hall conductivity metals. This non-trivial internal structure dramatically enriches the dynamics of plasmon propagation, enabling plasmon wavepackets to acquire geometric phases as they scatter. Strikingly, at boundaries these phases accumulate allowing plasmon waves that reflect off to experience a non-reciprocal parallel shift 
along the boundary displacing the incident and reflected plasmon trajectories. This plasmon Hall shift, tunable by Hall conductivity as well as plasmon wavelength, displays the chirality of the plasmon's current distribution and can be probed by near field photonics techniques. Anomalous plasmon dynamics provide a real-space window into the inner structure of plasmon bands, as well as new means for directing plasmonic beams. 
\end{abstract}

\pacs{pacs}
\maketitle

The internal structure of quasiparticles~\cite{Kane,Konig,XiaoValley,Onoda,bliokh04,XiaoLayer}, e.g., spin, valley degrees of freedom, while typically hidden from view can dramatically alter the behavior of particles when they are coupled with kinematic variables. A key ingredient for non-trivial band topology and associated gapless edge states in fermionic~\cite{Kane,Konig} and bosonic systems~\cite{Haldane,ZWang}, this coupling also warps bulk band {\it geometry} allowing the bands to acquire non-trivial (pseudo)-spin texture. This emergent texture enriches the dynamics of bulk band quasi-particles enabling them to acquire geometric phases that skew their trajectories and transform their response to external fields~\cite{OnodaGeo,Chang,Beenakker,XiaoBerry,Bliokh}.

The collective modes of an electron liquid -- plasmons -- comprise self-sustained oscillations of electric field $\vec E$ and charge density degrees of freedom. In two-dimensional (2D) metals, $\vec E$ and charge density oscillate in lock-step allowing deep sub-wavelength plasmons to be excited~\cite{NChen,ZFei}. These feature longitudinal electric (LE) modes (with field orientation and propagation direction that are aligned) that enable a range of nano-photonic properties far below the diffraction limit. As such, intense efforts have focussed on unraveling its dispersion and exploiting its tightly bound electric fields.

\begin{figure}[h!] \includegraphics[scale=0.64]{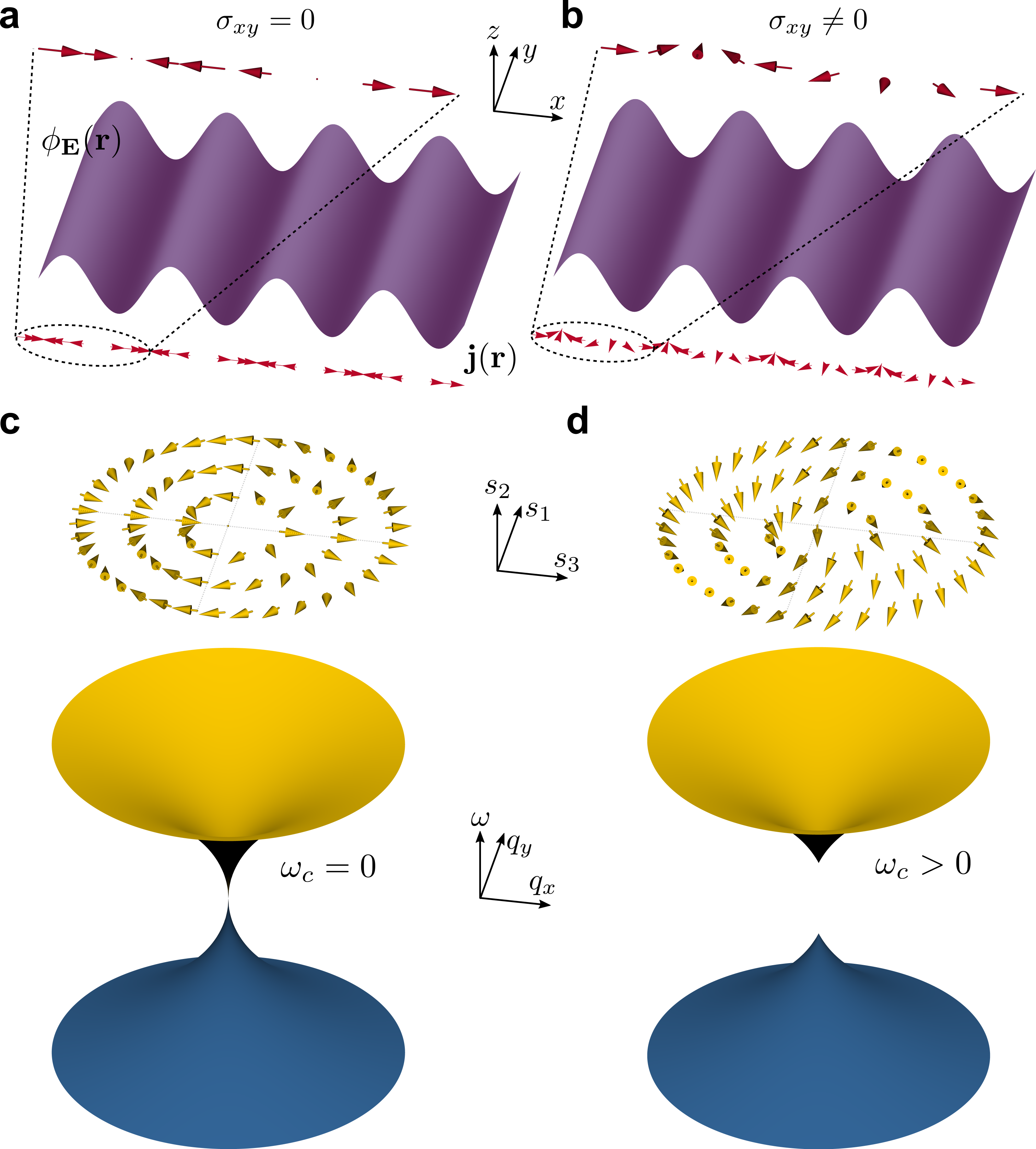}
\caption{ {\it Plasmon dispersion and internal current density texture}. 
({\bf a},{\bf b}) Current density (vector) fields $\vec j (\vec r)$ configuration for plasmons with $\sigma_{xy} = 0$ and $\sigma_{xy} \neq 0$ respectively. Note that while 
electric potential $\phi_{\vec E}(\vec r)$ take on the same longitudinal electric profile, $\vec j (\vec r)$ adopts a completely different texture. Plasmon dispersion for ({\bf c}) conventional plasmon ($\omega_c =0$), and ({\bf d}) magnetoplasmon modes ($\omega_c \neq 0$) in a two-dimensional metal (Eq.~\ref{eq:dispersion}), both in the non-retarded limit. Current density pseudo-spinor (Eq.~\ref{eq:pseoduspin}) for $\omega_c > 0$ exhibit a non-trivial canted texture; when $\omega_c =0$, pseudo-spinor only winds in plane. Blue band is complex-conjugate copy of primary plasmon band (see text).} 
\label{fig1} \end{figure}

Here we argue that in addition to $\vec E$ and charge density fields, deep sub-wavelength plasmons can possess a finer internal structure (Fig.~\ref{fig1}) that dramatically alters its dynamics. In particular, we find that plasmon current density configuration depends intimately on a metal's hall conductivity $\sigma_{xy}$, exhibiting an intricate pattern that cants away from the propagation direction when $\sigma_{xy}\neq 0$ (Fig.~\ref{fig1}b). This pattern is characterized by a plasmon current density pseudo-spin that tracks in-plane current density orientation and phase (Fig.~\ref{fig1}c,d), and enables a striking microscopic distinction between plasmons in conventional $\sigma_{xy} = 0$ metals and $\sigma_{xy} \neq 0$ metals. Plasmons in the latter possess a non-trivial hedgehog-like current density texture (Fig.~\ref{fig1}d) hosted within the 2D metal.

Crucially, the non-trivial current density pseudo-spinor texture directly affects the dynamics of plasmon waves, allowing them to accumulate geometric phases as they propagate. In particular, we demonstrate that for $\sigma_{xy} \neq 0$, plasmon plane waves acquire an additional geometric phase (directly tied to the non-trivial texture) when they scatter off a boundary. Strikingly, these phases accrue when plane waves are superposed, allowing wavepackets that reflect off boundaries to experience a parallel translation along the boundary -- plasmon Hall shift [Fig.~\ref{fig3}] --  displacing the incident and reflected plasmon beams by multiple plasmon wavelengths. As we show below, this plasmon Hall shift is geometrical and directly arises from a geometric connection between the incident and reflected waves. As a result, a non-reciprocal plasmon Hall shift manifests even though plasmons are charge neutral, with sign controlled by the incident wavevector and $\sigma_{xy}$ (vanishing when $\sigma_{xy} = 0$).  

We note that electric field and charge density configuration of deep subwavelength plasmons are directly addressable by spectroscopic probes; these also determine the  plasmon's energy spectrum. In contrast, local current density patterns are typically obscured from view, and are difficult to detect directly via conventional probes. 
Nevertheless, the unusual plasmon dynamics we discuss here manifest directly because of the current density texture that is {\it hidden} in the motion of electrons in the 2D metal plane. Just as the inner degrees of freedom of electrons can alter electron motion~\cite{Chang,Beenakker,XiaoBerry}, the plasmon's current density configuration acts as an internal structure of the plasmon quasiparticle and enables unconventional dynamics.

We expect non-trivial plasmon textures to manifest in 2D metals with non-zero hall conductivity, including conventional two dimensional electron gas with an applied magnetic field, as well as anomalous Hall metals~\cite{Song,vignale,low}. Further, the exposed 2D surface of these materials are a particularly ripe venue to observe the associated plasmon Hall shift and can be directly probed by scanning near-field scattering microscopy techniques~\cite{NChen,ZFei} recently developed to map out plasmon trajectories. These can provide a sensitive window into the internal structure of plasmon quasiparticles, as well as new ways for routing plasmons in the extreme sub-wavelength regime. 

\vspace{2mm}
{\it Plasmon current texture and pseudospin ---} To analyze the geometric properties of plasmonic bands, we begin by examining its associated dynamical electric fields close to a 
2D metal confined in the plane (at $z=0$). 
The dynamical evolution of electric fields ${\boldsymbol {\mathcal E}} (\vec r, z, t)$ obey Maxwell's equations so that
\be
{\boldsymbol \nabla}  \times [{\boldsymbol \nabla} \times {\boldsymbol {\mathcal E}} (\vec r, z, t)]  
 = 
  \frac{\omega^2}{c^2}  {\boldsymbol {\mathcal E}} (\vec r, z, t) - i \frac{4\pi \omega}{c^2} {\boldsymbol {\mathcal J}} (\vec r, z, t) ,
\label{eq:FAgeneral}
\ee
where $\vec r=(x,y)$ lie in the 2D plane, the current density 
$
{\boldsymbol {\mathcal J}} (\vec r, z, t) = ( j_x , j_y , 0) \exp ( i \omega t - i \vec q \cdot \vec r)  \delta(z)   ,
$
is confined to the 2D metal, and $\vec j = (j_x, j_y)$ is the amplitude of the 2D current density. We have set the relative electric/magnetic permittivity/permeability to unity for simplicity.

Away from the 2D plane ($z \neq 0$), 
${\boldsymbol {\mathcal J}} (\vec r, z, t)$ vanishes yielding electric fields that obey the familiar profile $ {\boldsymbol {\mathcal E}} (\vec r, z, t) = \vec E \exp ( i \omega t - i \vec q \cdot \vec r - i k_z z) $ with $q_x^2+q_y^2 + k_z^2 = \omega^2 / c^2$, where $\vec E$ is a constant vector that denotes the polarization in the 2D plane, and $\vec q = (q_x, q_y)$. Importantly, the free charge currents induced in the 2D plane enable tightly-bound electromagnetic solutions close to the metallic plane~\cite{Chiu}. In these plasmonic modes, $k_z^2 < 0 $ so that the electric field amplitude decays exponentially away from $z=0$ so that 
${\boldsymbol {\mathcal E}}^\pm (\vec r, z, t) = \vec E^\pm \exp ( \mp \beta z) \exp ( i \omega t - i \vec q \cdot \vec r)$, where $\beta = ( q^2 - \omega^2/c^2)^{1/2} >0$, and $q^2 = q_x^2 + q_y^2$. The $\pm$ superscripts denote the regions $z>0$ or $z < 0$, respectively. 

Substituting the form of the tightly-bound fields into Eq.~\ref{eq:FAgeneral} we obtain electric fields in the 2D plane as
\be
\lp \begin{array}{c}
E_x \\
E_y
\end{array} \rp
 = {\boldsymbol {\mathcal F}} 
\lp \begin{array}{c}
j_x \\
j_y
\end{array} \rp, \quad {\boldsymbol {\mathcal F}}  = \frac{2\pi i}{\omega \beta }
\lp \begin{array}{cc}
\beta^2 - q_y^2 & q_x q_y \\
 q_x q_y & \beta^2 - q_x^2
\end{array} \rp    .
\label{eq:EJmaxwell}
\ee
Plasmons (collective modes of the electron liquid) are obtained by the self-sustained solutions of Eq.~\ref{eq:EJmaxwell} and the constitutive relations governing the transport of free carriers in the 2D plane, namely its conductivity tensor: $ \vec j = \boldsymbol{\sigma} \vec E$. Eliminating $\vec E$ in favor of the current density $\vec j$, we obtain 2D plasmons from solutions of 
\be
\boldsymbol{\mathcal{M}} \, \vec j = 0, \quad \boldsymbol{\mathcal{M}} = \boldsymbol{\mathcal{F}} - \boldsymbol{\sigma}^{-1},
\label{eq:m}
\ee
where the constitutive relation (encoded in $\boldsymbol{\sigma}$) determines both the dispersion and the configuration of current density in the 2D metal. As we display below, the current density configuration can wind as a function of plasmon propagation direction and wavelength leading to non-trivial geometrical effects. 

We first consider plasmons in a conventional 2D metal with applied magnetic field $\vec B = B_0 \hat{\vec{z}}$. Its in-plane conductivity is described by the Drude model:  
\begin{align}
\sigma_{xx} = \frac{(1 + i \omega \tau)~\sigma_0}{(1+ i \omega \tau)^2 + (\omega_c \tau)^2}    , \quad
\sigma_{xy} = \frac{ - \omega_c \tau~\sigma_0}{(1 + i \omega \tau)^2 + (\omega_c \tau)^2},    
\label{eq:sigma}
\end{align}
where $\sigma_0 = D_0 \tau$ is the static conductivity with $D_0 = n e^2/m$ the Drude weight, $\tau$ is the scattering time, $m$ is the effective mass of the electron, and $\omega_c = e B_0 / m$ is the cyclotron frequency.

Taking the non-retarded limit ($ q \gg \omega/c$) and analyzing the collisionless regime wherein $\tau \gg 1/\omega$, we solve Eq.~\ref{eq:m} using Eq.~\ref{eq:sigma} to obtain the 2D bulk plasmon dispersion
\be
\omega = \pm \omega_{q}, \quad  \omega_{q} =  \sqrt{a_0 q  + \omega_c^2}   .
\label{eq:dispersion}
\ee
where $a_0= 2 \pi D_0 $. When $\omega_c = 0$, $\omega_q$ yields to the familiar 2D plasmon dispersion (see Fig.~\ref{fig1}c). $\omega_c \neq 0$ corresponds to (gapped) bulk 2D magnetoplasmon spectrum (see Fig.~\ref{fig1}d)~\cite{Chiu}. 
Crucially, we find that the corresponding eigenmodes of Eq.~\ref{eq:m} in the $x$-$y$ basis are
\be
\vec u^\pm (\vec q) 
= \lb \begin{array}{c} j_x (\vec q)\\ j_y (\vec q) \end{array} \rb^\pm
= \frac{\mathcal{N}}{q}
\lp \begin{array}{c}
  \mp i q_x + \eta_q q_y \\
  \mp i q_y - \eta_q q_x
\end{array} \rp.
\label{eq:eigenJ}
\ee
Here $\eta_q = \omega_c/\omega_q \in [-1,1]$ is a dimensionless parameter that captures the relative magnetic field strength, and $\mathcal{N} = (1+\eta_q^2)^{-1/2}$ is a normalization constant: $\la \vec u^\pm (\vec q) | \vec u^\pm (\vec q) \ra = 1$.
We emphasize that $\eta_q$ represents the Hall coupling between the current $j_{x(y)}$ with its perpendicular electric field $E_{y(x)}$. 
We note, parenthetically, that the eigenmodes in each of the bands, $\vec u^+ (\vec q)$ and $\vec u^- (\vec q)$,
are locked to each other: $[\vec u^+ (\vec q)]^* = - \vec u^- (-\vec q)$.
This ensures that fields governed by Maxwell's equations are real-valued (see Supplementary Information, {\bf SI}~\cite{supp}). Here and in the following, we focus on $\vec u (\vec q) = \vec u^+ (\vec q)$ and drop the band index $+$ for brevity. 

Canting of current directions (see Eq.~\ref{eq:eigenJ}) enable the plasmons to exhibit a chirality and acquire non-trivial phases as they propagate. To illustrate this, we analyze the pseudospin of the plasmon eigenmodes, $s_i (\vec q) =  \la \vec u (\vec q) | \sigma_i | \vec u (\vec q) \ra$: 
\be
s_1 = \frac{1- \eta_q^2 }{1+ \eta_q^2 } \sin 2\phi_s , 
\quad  s_2 = \frac{- 2\eta_q}{1+\eta_q^2}  ,\quad s_3 =  \frac{1-\eta_q^2}{1+\eta_q^2} \cos 2\phi_s ,  
\label{eq:pseoduspin}
\ee
where $\sigma_i$ are Pauli matrices ($i=1,2,3$), and ${\rm tan} \phi_s = q_y/q_x$. As shown in Fig.~(\ref{fig1}{\bf a}), when $\omega_c=0$, the conventional 2D metal plasmons possesses a pseudospin $\vec s$ components in the plane ($s_2=0$, $s_1,s_3 \neq 0$) and winds twice as $\phi_s$ is varied from $0$ to $2\pi$. In contrast, when $\omega_c \neq 0$, the pseudospin $\vec s$ cants out of the plane, see Fig.~(\ref{fig1}{\bf b},{\bf d}). While the pseudospin continues to wind twice as $\phi_s$ varies from $0 \to 2\pi$, its $s_2$ component changes pitch, leading to a topologically non-trivial pseudospin texture. As we explain below, this hedgehog-like pseudo-spin texture enables plasmons to acquire non-trivial geometric phases as they propagate.

We note that in contrast to the non-trivial current density texture, electric field ${\boldsymbol \varepsilon} (\vec q)$ directionality is insensitive to Hall conductivity and have orientation
\be
{\boldsymbol \varepsilon} (\vec q) = {\boldsymbol {\mathcal F}}(\vec q)\vec u (\vec q) = (2 \pi \mathcal{N}/ \omega_q) \vec q  ,
\label{eq:eigenE}
\ee 
that continues to point along $\vec q$. As a result, the direction of ${\boldsymbol \varepsilon} (\vec q)$ is the same for both plasmons at zero magnetic field, 
and plasmons with a finite applied magnetic field; 
${\boldsymbol \varepsilon} (\vec q)$ plasmons are longitudinal electric (LE) modes for all $\eta_q$.

\begin{figure}[t!] \includegraphics[scale=0.7]{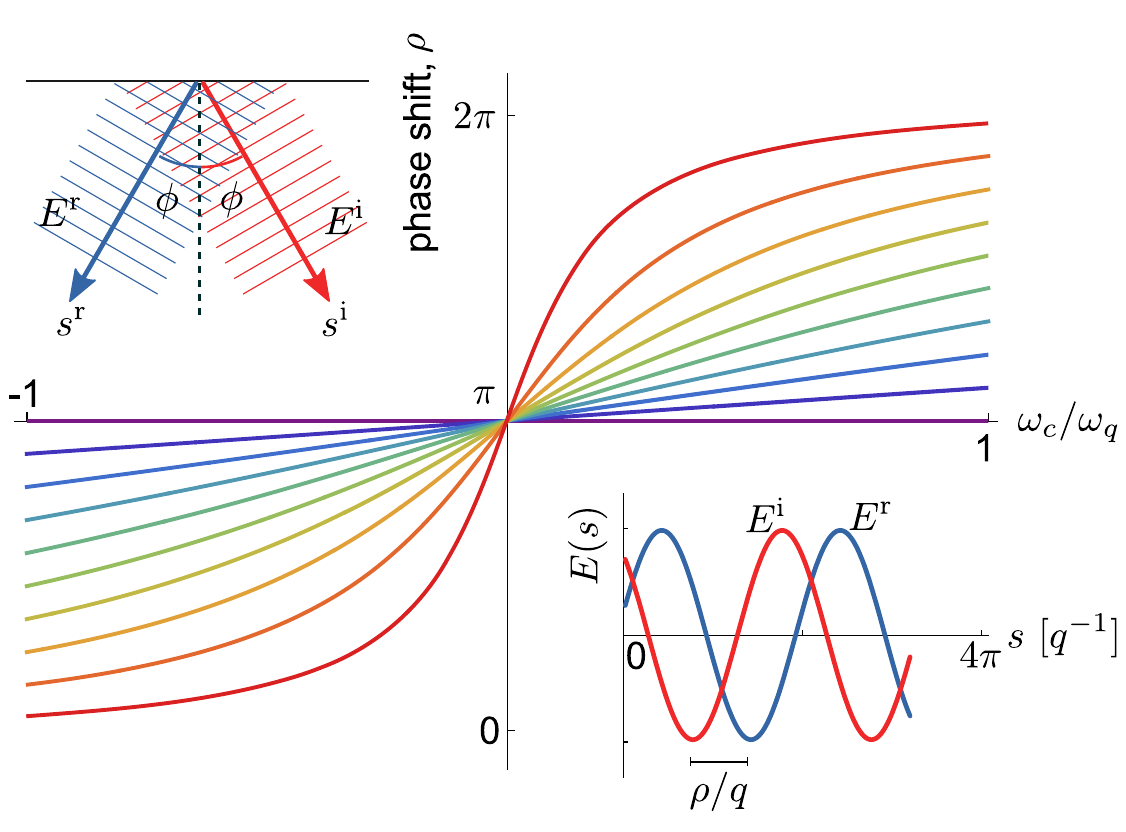}
\caption{ {\it Plasmon geometric phase.} Phase shift between incident and reflected plane waves scattered off a boundary at different incident angles and relative magnetic field strength obtained from Eq.~\ref{eq:rho}; while geometric phase shifts appear in both the current density and electric field, here we show phase shift for (longitudinal) electric field along the directions $-\vec q^{\rm i}$ ($\vec q^{\rm r}$) away from the interface. Lines from purple (closest to the $x$-axis) to red (furthest away from the $x$-axis) correspond to incident angles $\phi_n =(\pi / 2)  (n / 10)$ with $n$ integers ranging from $n\in \{0, \cdots, 9\}$. (Upper inset) Schematic for the incident and reflected plane wave. (Lower inset) Electric field $E^{\rm i}$ ($E^{\rm r}$) of the incident (reflected) plane wave exhibit a non-trivial shift $\rho$ tunable by Hall conductivity (magnetic field).}  
\label{fig2} \end{figure}

\begin{figure}[t!] \includegraphics[width=1\columnwidth]{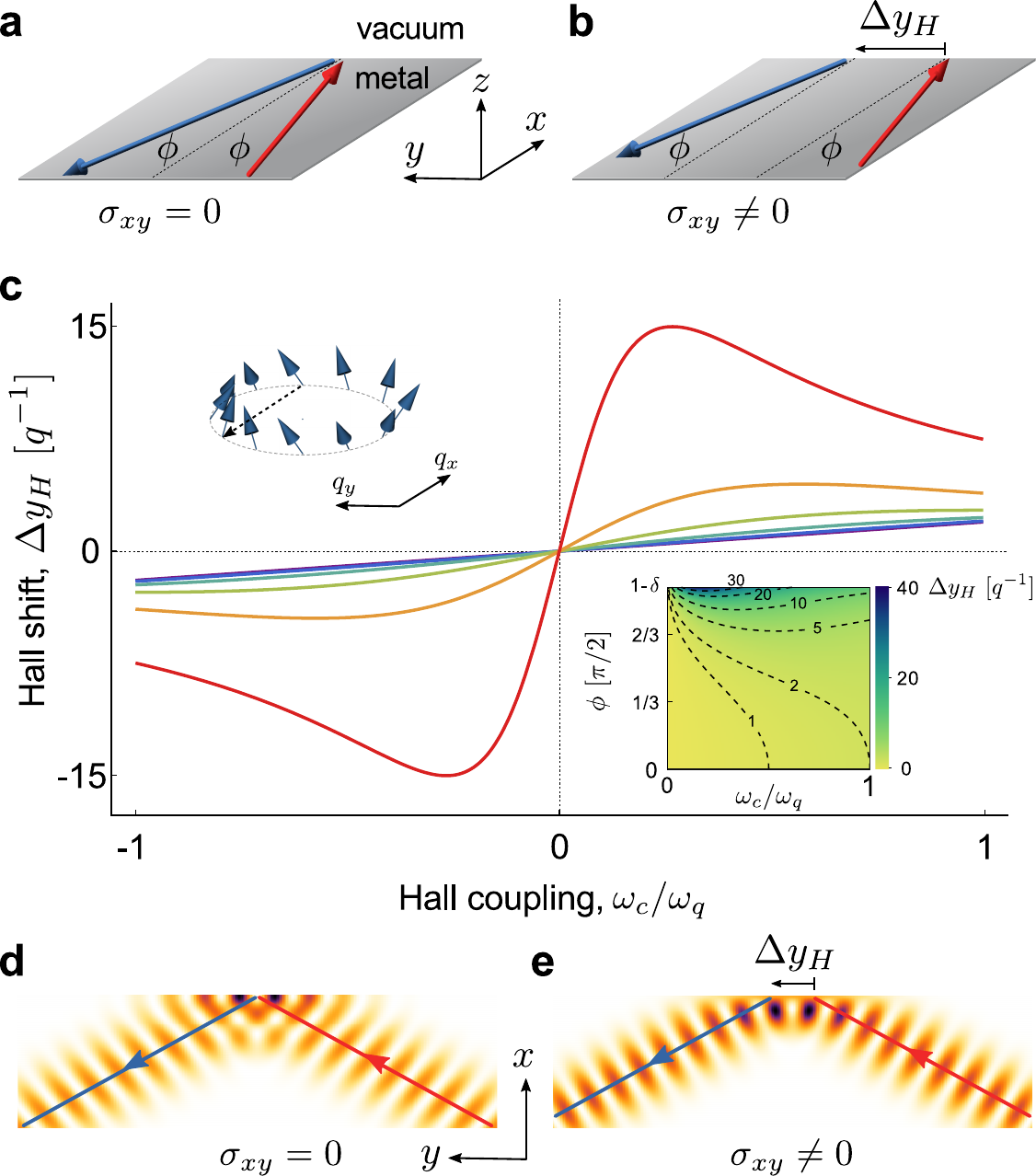}
\caption{{\it Plasmon Hall shift}. 
({\bf a},{\bf b}) Schematic of plasmon Hall shift between the incident and reflected wavepackets with incident angle $\phi$. The Hall shift $\Delta y_H$ is non-zero when $\sigma_{xy} \neq 0$. ({\bf c}) Plasmon Hall shift for 2D magnetoplasmons (in units of inverse wavevector $q^{-1}$) from Eq.~\ref{eq:hallshift} at various incident angles $\phi$ can be tuned by the Hall coupling $\eta = \omega_c/\omega_q$.  Lines from purple (closest to the line $\Delta y_H = 0$) to red (furthest away from the line $\Delta y_H = 0$) correspond to incident angles $\phi_n =  (\pi/2) (n / 6) $ with $n$ integers ranging from $n \in \{0, \dots, 5  \}$.
(Upper inset) displays the pseuodo-spinor orientation on a bulk plasmon iso-frequency contour highlighting the different orientations between incident and reflected waves. (Lower inset) Contour plot of Hall shift as a function of Hall coupling and $\phi$. We have used $\delta = \pi/20$. ({\bf d},{\bf e}) Numerically computed electric field intensity $[ \Re \vec E^{\rm i} (\vec r) + \Re \vec E^{\rm r} (\vec r) ]^2 $ from Eq.~\ref{eq:E}, without and with magnetic field (at $\eta_q = 1$ limit), for an incident wave with a Gaussian distribution and fixed energy. The width of the wave is $\pi / q$ and the incident angle is $\pi/3$, with a Hall shift $4/q$. Red and blue lines are beam trajectories obtained from Eq.~\ref{eq:center}.
} \label{fig3} \end{figure}

\vspace{2mm}
{\it Plasmon geometric phase and boundary conditions --- } 
We now illustrate the effect of the canted current density configuration in the dynamics of bulk plasmons.
We consider a semi-infinite 2D metal (boundary at $x=0$) (see Fig.~\ref{fig2} and \ref{fig3}a,b); magnetic field is applied along $\hat{\vec{z}}$. 
Plasmon waves that impinge on $x=0$ will get reflected elastically. In the main text, we will focus on the regime of total reflection, see {\bf SI}~\cite{supp} for a more general discussion. To proceed, we construct the incident (reflected) plasmon wavefronts impinging on (moving away from) the boundary. The current density distribution is a linear superposition of eigenmodes $\vec u (\vec q)$:
\begin{align}
\vec J^{\rm i,r}  (\vec r )   =  \sum_n g_n^{\rm i,r}  \vec u (\vec q_n^{\rm i,r})
\exp (- i \vec q_n^{\rm i,r} \cdot \vec r )
\label{eq:incidentwave}
\end{align}
where $g_n^{\rm i,r}$
is a complex number that captures the current density amplitude and relative phases between components in the linear superposition.

First, we note that $\vec q_n^{\rm i,r}$ for each component $n$ are related by translational symmetry along $y$ and energy conservation yielding  $ (q_{n,x}^{\rm i}, q_{n,y}^{\rm i}) = (- q_{n,x}^{\rm r}, q_{n,y}^{\rm r}) $. Importantly, ${\vec J}^{\rm i,r} $ satisfies a current boundary condition along $x$ so that at the boundary the current vanishes: 
\be
\hat{\vec{x}} \cdot [\vec J^{\rm i} (x=0, y) + \vec J^{\rm r} (x=0, y)] = 0.
\label{eq:BC1}
\ee
In order to satisfy Eq.~\ref{eq:BC1} the complex superposition factors for incident and reflected waves obey the relation
\be
\frac{g^{\rm r}_n } { g^{\rm i}_n } = - \frac{ \hat{\vec{x}} \cdot \vec u  (\vec q_n^{\rm i} ) }{ \hat{\vec{x}} \cdot  \vec u (\vec q_n^{\rm r})  }=  \exp ( i \rho_n  ), 
\label{eq:rho}
\ee
where $\rho_n$ is a real number representing the phase shift between $g_n^{\rm r}$ and $g_n^{\rm i}$, because the norm of $- u_x  (\vec q_n^{\rm i} ) /  u_x (\vec q_n^{\rm r})$ is unity. We note that when $\eta_q = 0$ (zero magnetic field), Eq.~\ref{eq:rho} yields the familiar trivial phase $\rho_n = \pi$. However, when $\eta_q \neq 0$ (finite magnetic field), $\rho_n \neq \pi $ and picking up a non-trivial phase. Importantly, since the phase shifts arise from the reflection coefficients, non-trivial $\rho_n$ manifest in both current density and electric field profiles. As a demonstration, we plot the phase shift $\rho$ in Fig.~\ref{fig2} for a plane wave scattering off the boundary. In the insets of Fig.~\ref{fig2}, we have displayed the electric field profile obtained by applying Eq.~\ref{eq:eigenE} to Eq.~\ref{eq:incidentwave}. These exhibit non-trivial phases between incident $E_i$ and reflected $E_r$ waves when $\eta_q\neq 0$ (right inset).  

We emphasize that $\rho_n$ is directly related to geometry of the canted pseudo-spin structure (Fig.~\ref{fig1}). Writing $\rho_n \rightarrow \rho (\vec q)$ as a continuous variable of incident wavevector $\vec q$, we find that the nontrivial phase $\rho (\vec q)$ is intimately linked to a $\vec q$-dependent geometric connection:
\be
{\boldsymbol {\cal A} } (\vec q, \hat {\vec n} )= 
 \la u_{\hat {\vec n}} (\vec q) | i {\boldsymbol \nabla}_{\vec q} |  u_{\hat {\vec n}} (\vec q) \ra,
\label{eq:Ageneral}
\ee
via ${\boldsymbol \nabla}_{\vec q} \rho (\vec q) = { \boldsymbol {\cal A} } (\vec q_0^{\rm r},\hat {\vec n} ) - { \boldsymbol {\cal A} } (\vec q_0^{\rm i}, \hat {\vec n} )$, where $| u_{\hat {\vec n} } (\vec q) \ra =  | u (\vec q) \cdot \hat {\vec n} \ra \la \hat {\vec n} \cdot u (\vec q)   | u (\vec q) \cdot \hat {\vec n} \ra^{-1/2} $, and $\hat {\vec n} $ is a unit vector denoting the direction of a interface [for $\rho (\vec q)$ defined by the boundary; in Eq.~\ref{eq:rho}, ${\hat {\vec n} }$ is $\hat {x}$]. We note that $\vec u (\vec q)$ are classical eigenmodes and {\it do not} possess gauge freedom. As a result, the corresponding geometric connection ${\boldsymbol {\cal A} } (\vec q, \hat {\vec n})$ is fixed in each system (characterized by its conductivity tensor, $\boldsymbol \sigma$), and is real
\footnote{The realness of ${\boldsymbol {\cal A} } (\vec q, \hat {\vec n})$ can be discerned as follows. First, we note that $ {\boldsymbol \nabla}_{\vec q} [\la u_{\hat {\vec n}} (\vec q) |  u_{\hat {\vec n}} (\vec q) \ra] =  [\la { u_{\hat {\vec n}} (\vec q) | \boldsymbol \nabla}_{\vec q}  u_{\hat {\vec n}} (\vec q) \ra]^*  + \la { u_{\hat {\vec n}} (\vec q) | \boldsymbol \nabla}_{\vec q}  u_{\hat {\vec n}} (\vec q) \ra$. Next, we note that since ${\boldsymbol \nabla}_{\vec q} [\la u_{\hat {\vec n}} (\vec q) |  u_{\hat {\vec n}} (\vec q) \ra]$ must also vanish, $ \la { u_{\hat {\vec n}} (\vec q) | \boldsymbol \nabla}_{\vec q}  u_{\hat {\vec n}} (\vec q) \ra$ must be pure imaginary. As a result,  ${\boldsymbol {\cal A} } (\vec q, \hat {\vec n}) =  \la u_{\hat {\vec n}} (\vec q) | i {\boldsymbol \nabla}_{\vec q} |  u_{\hat {\vec n}} (\vec q) \ra$ is pure real. This is different from the the typical Berry connection, which defines the parallel transport in local tangent plane spanned by $| \vec u (\vec q) \ra$. Here ${\boldsymbol {\cal A} } (\vec q, \hat {\vec n})$ is a geometric connection defining a parallel transport projected to a fixed plane determined by the reflecting boundary.}.

\vspace{2mm}
{\it Plasmon Hall shift --- } As we now show, the non-trivial phases acquired above can build-up to dramatic effect for wavepackets (superposition of plane waves). In particular, wavepackets reflected off a boundary, experience a real space shift (see Fig. \ref{fig3}b) with the reflected wavepacket center translated along the boundary when a magnetic field is applied (relative to the zero magnetic field case): the plasmon Hall shift. 

To clearly illustrate the plasmon Hall shift, we analyze wavepackets of electric field strength $\vec E(\vec r)$ (see {\bf SI}~\cite{supp} for other types of wavepackets). Since the intensity distribution $|\vec E (\vec r)|^2$ can be directly measured by scanning near-field optical probes~\cite{NChen,ZFei}, we will focus on $|\vec E^{\rm i,r}(\vec r)|^2$ peak position to track the location of the plasmon wavepacket in real space. $\vec E(\vec r)$ plasmon wavepacket can be directly obtained from Eq.~\ref{eq:incidentwave} via Eq.~\ref{eq:eigenE}. In constructing the wavepacket we will take a continuous superposition of eigenwaves so that the incident complex superposition factors $g_n^i \to f(\vec q) = |f(\vec q)| \exp[i\delta(\vec q)]$ where $|f(\vec q)|$ is real with a narrow and single center at $\vec q_0 = (q_0 \cos \phi_0, q_0 \sin \phi_0)$ with $\phi_0$ the incident angle, and $\delta (\vec q)$ is the phase distribution profile. We obtain
\begin{align}
{\vec E}^{\rm i,r} (\vec r)  = \int {\rm d} \vec q f (\vec q)  p^{\rm i,r}(\vec q) {\boldsymbol \varepsilon}^{\rm i,r} (\vec q ) 
e^{ \mp i q_x x - i q_y y } ,
\label{eq:E}
\end{align}
where ${\boldsymbol \varepsilon}^{\rm i} (\vec q ) = {\boldsymbol \varepsilon} ( q_x, q_y )$ corresponds to the incident wave, ${\boldsymbol \varepsilon}^{\rm r} (\vec q ) = {\boldsymbol \varepsilon} (-q_x,q_y)$ corresponds to the reflected the wave, $p^{\rm i}(\vec q) = 1$, and the reflection coefficient $p^{\rm r}(\vec q)  = \exp [i \rho (\vec q) ]$. 

Using Eq.~\ref{eq:E}, we track the center of the electric field intensity $[\Re {\vec E}^{\rm i,r} (\vec r) ]^2$. We determine the center of the narrowly distributed wavepacket (beam)  using the standard method of stationary phases (see ``Methods''), obtaining the trajectories for the center of the incident $y_c^{\rm i} (x)$ and reflected $y_c^{\rm r} (x) $ beams as 
\be
y_c^{\rm i} (x)= - \kappa_0 x +  y_0  ,\quad y_c^{\rm r} (x)=  \kappa_0 x + y_0 + \Delta y_H , 
\label{eq:center}
\ee
where $\kappa_0 = - \tan \phi_0 $. 
While $y_0 = [{\rm d} \delta (\vec q) /{\rm d} q_y ]_{\vec q = \vec q_0}  $ denotes a reference point (e.g., absolute position of wavepacket along $y$) fixed by the relative phase factors of complex superposition factors in the incident wavepacket.

Importantly, the incident and reflected trajectories, $\quad y_c^{\rm i,r}$, are displaced at $x=0$ by $\Delta y_H$. This Hall shift arises from the non-trivial geometric phases (Eq.~\ref{eq:rho}) sustained by each $\vec q$-component  of the wavepacket as it is reflected off the boundary. Since phases are slightly mismatched for different $\vec q$, their phase gradient produces: 
\begin{align} 
\Delta y_H & =  [  {\rm d} \rho (\vec q + \vec k)/ {\rm d} k_y  ]_{\vec q = \vec q_0, \vec k=0}  \nn
 & =  \kappa_0 [ { \boldsymbol {\cal A} } (\vec q_0^{\rm r}, \hat {x}) - { \boldsymbol {\cal A} } (\vec q_0^{\rm i}, \hat {x}) ]_x + [ { \boldsymbol {\cal A} } (\vec q_0^{\rm r}, \hat {x}) - { \boldsymbol {\cal A} } (\vec q_0^{\rm i}, \hat {x}) ]_y  ,
\label{eq:deltay} 
\end{align}
where we have used Eq.~\ref{eq:rho} and Eq.~\ref{eq:Ageneral}. Eq.~\ref{eq:deltay} shows that the real space shift is directly related to the geometric properties of the plasmon band, allowing direct access into its internal structure via real space mapping.

Using the form of the current density $u_x$ in Eq.~\ref{eq:eigenJ} and Eq.~\ref{eq:Ageneral} we find the Hall shift for 2D magnetoplasmons as
\begin{align}
\Delta y_H^{\rm mp} 
 =  \frac{ 1 }{q_0}
 \frac{2 \eta_0 \sec \phi_0 }{ \cos^2 \phi_0 + \eta_0^2 \sin^2 \phi_0 }    ,
\label{eq:hallshift}
\end{align}
where $\eta_0 = \omega_c/\omega_{q_0}$. An explicit calculation of $\Delta y_H^{\rm mp}$ using Gaussian wavepackets reproduces Eq.~\ref{eq:hallshift}, see {\bf SI}~\cite{supp}. 
We note that $\Delta y_H^{\rm mp} $ (Fig.~\ref{fig3}c) is directly proportional to $\eta_0$, vanishing in the presence of TRS (e.g., at $\omega_c = 0$). Further, it exhibits a sharp $\phi$ dependence with $\Delta y_H^{\rm mp} $ allowing the effect to be further tuned. 

Curiously, although plasmons are charge neutral excitations, they can still experience a Hall shift similar to that of charged particles. Charged particles experience a transverse flow that depends on the sign of the charge, applied electric field, and magnetic field direction. The direction of the plasmon Hall shift above is determined by the Hall coupling (set by the Hall conductivity), and the direction of the wave vector perpendicular to the boundary $q_x$ giving: ${\rm sgn} (\Delta y_H) = \sgn (\eta_q q_x)$. As a result, $\Delta y_H$ is even in $\phi$, yielding the {\it same} direction of shift for both positive and negative $\phi$ plasmon waves incident on a single boundary. The non-reciprocal shift of bulk plasmon wavepackets may provide a novel way of filtering plasmons that propagate in opposite directions in $q_y$ (along the boundary) but with the same sign in $q_x$ (perpendicular to the boundary). 

\vspace{2mm}
{\it Anomalous plasmon Hall shift--- } While we focussed on conventional 2D metals at finite magnetic field above, a non-trivial plasmon band geometry can also be achieved at zero field. For example, plasmon excitations can be supported in anomalous Hall metals~\cite{Song}, electrons in strained graphene or in proximity to a skyrmion lattice~\cite{vignale}, as well as out-of-equilibrium systems~\cite{Song,low} all of which possess a Hall conductivity in the absence of an applied magnetic field.

To illustrate the anomalous plasmon Hall shift, we concentrate on the first example: anomalous Hall metals. We take a simple phenomenological model for the conductivity tensor with longitudinal component $\sigma_{xx}  = \sigma_0/(1+ i\omega \tau)$ in the usual Drude form, and Hall conductivity $\sigma_{\rm AH}$ (valid for small frequencies and wavevectors). Following the same procedure in Eqs. \ref{eq:FAgeneral}-\ref{eq:m} above, we obtain a gapless bulk plasmon frequency as $\omega_q^{\rm A} =  \sqrt{a_0 q} $, where $a_0 = 2 \pi D_0$. Furthermore, just as conventional Hall conductivity (from a Lorentz force) skew the current density orientation (see Fig.~\ref{fig1}), anomalous Hall conductivity also yields a canted current density orientation. The current density orientation for anomalous Hall plasmon eigenmodes map precisely to Eq.~\ref{eq:eigenJ} but with $\eta_q^{\rm A} =   \omega_q^{\rm A} \sigma_{\rm AH} / D_0$.

As a result of the non-trivial bulk plasmon current density texture, bulk plasmons in anomalous Hall metals also experience a plasmon Hall shift. Indeed, applying the eigenmodes obtained above, and considering the same geometry as in Fig.~\ref{fig3}, we obtain the Hall shift
\be
 \Delta y_H^{\rm A} = \frac{ 1 }{q_0}
 \frac{2  \eta_0^{\rm A}  \sec \phi_0 }{ \cos^2 \phi_0 +  ( \eta_0^{\rm A} )^2 \sin^2 \phi_0 }  ,
\label{eq:hallshiftA}
\ee 
that is controlled by incident wave vector $\vec q_0 = (q_0 \cos \phi_0, q_0 \sin \phi_0 )$, and dimensionless Hall coupling $\eta_0^{\rm A} = \omega_{q_0}^{\rm A} \sigma_{\rm AH} / D_0$. In contrast to the case of magnetoplasmons above, we note the Hall coupling $\eta_0^{\rm A}$ can be larger than unity. As a result, anomalous Hall plasmons may display a larger Hall shift. 

\vspace{2mm}
{\it Discussion -- }
The plasmon Hall shifts we discuss here can be probed by a variety of experimental techniques. For example, a focussed plasmon wavepacket can be launched
at a reflecting boundary (e.g., via designer nano-anttenas~\cite{Alonso} fixed on the sample). The subsequently reflected plasmon wavepacket can be tracked
by out-coupling light using scanning near-field tip and microscope~\cite{ZFei,NChen}. Since the Hall shift is tunable by both incident angle and wavevector $\vec q$, as well as an applied magnetic field, the reflected plasmon trajectory can enable an optical mapping of the inner structure of bulk plasmon bands, namely its current density orientation/texture. This is similar to the role spin-resolved photoemission spectroscopy plays in mapping the spinor structure of topologically non-trivial electronic bandstructures 

We note that the plasmon Hall shift while reminiscent of the Hall effect of light~\cite{Onoda,bliokh04} is distinct from it
in both its phenomenology and microscopic origin. First, the plasmons in 2D metals we consider are deep in the non-retarded regime exhibiting a dramatically compressed wavelength than that of light. This renders the plasmon Hall shifts discussed above far below the diffraction limit; wavepacket shifts of light 
are typically on the order of the free space wavelength~\cite{Onoda,bliokh04}. Second, geometrical phases in optics and photonics (in the retarded limit) arise from a texture in the {\it electric} polarization of light~\cite{bliokhreview}. 
In contrast, deep in the non-retarded limit, plasmons are purely composed of LE modes (electric field aligned with $\vec q$) which possess identical orientations for $\sigma_{xy} = 0$ metals and $\sigma_{xy}\neq0$ metals. Non-trivial plasmon geometric phases instead arise from a texture of the {\it current density} characterizing the collective motion of electrons in the 2D metal plane; this current density texture is unique to plasmon modes in metals and can be viewed as a ``hidden'' internal degree of freedom hosted within the 2D metallic plane clearly delineating the behavior of plasmons in $\sigma_{xy} = 0$ metals and $\sigma_{xy}\neq0$ metals. 
Lastly, we note that geometrical phases acquired by plasmons enable their motion to be controlled by conductivity and a homogeneous applied magnetic field. This provides an on-demand means of tuning the dynamics of plasmon wavepackets.

In summary, the internal current density configuration of plasmons in a metal can be transformed by transverse Hall conductivity and as a result, enable a new set of plasmon kinematics. In particular, we find that $\sigma_{xy} \neq 0$ plasmon wavepackets reflected off a boundary can acquire a non-reciprocal Hall shift parallel to the boundary. While we have focused on the plasmon Hall shift here, we expect that non-trivial plasmon band geometry can also enable a wealth unconventional dynamics not available when $\sigma_{xy} = 0$. For example, geometric phases can be accumulated by plasmon wavepackets as they propagate through inhomogeneous media, and may lead to anomalous plasmon wavepacket dynamics (e.g., anomalous velocity) in much the same way as for wavepackets of electrons~\cite{Chang,Sundaram,XiaoBerry} and photons~\cite{Onoda,bliokh04,OnodaGeo}; it can also lead to a topological structure of plasmon bands~\cite{Dafei}. Non-trivial band geometry provides a new tool box in which to control the dynamics of plasmons, as well as a window through which to view the internal structure governing the nano-photonic properties of materials.

\begin{acknowledgments}
We thank Eddwi Hesky Hasdeo and Yidong Chong for useful discussions. We are grateful to Alessandro Principi for a critical reading. This work was supported by the Singapore National Research Foundation (NRF) under NRF fellowship award NRF-NRFF2016-05.
\end{acknowledgments}

\section{Methods}

{\it Wavepacket center and stationary phases -- } In this section, we detail the standard method of stationary phases used to determine the wavepacket center, and its trajectory. Using Eq.~\ref{eq:E} of the main text, the slowly varying envelope for electric field intensity $[ \Re {\vec E}^{\rm i,r} (\vec r) ]^2$ can be expressed as~\cite{supp}
\begin{align}
F^{\rm i,r} (\vec r)=& \int  {\rm d} \vec q {\rm d} \vec k~ f_E^{\rm i, r} (\vec q,\vec k)  \exp [ i \theta^{\rm i,r} (\vec q, \vec k, \vec r) ], 
\label{eq:E2}
\end{align}
where $f_E^{\rm i, r} (\vec q, \vec k) =  | f (\vec q)| |f (\vec q+ \vec k)|  
 {\boldsymbol \varepsilon}^{\rm i,r} (\vec q )  \cdot  {\boldsymbol \varepsilon}^{\rm i,r} (\vec q + \vec k) $ is a real function of $(\vec q, \vec k)$, and
\begin{align}
\theta^{\rm i}  &= \Delta \delta (\vec q, \vec k)  - k_x x - k_y y, \nonumber \\
\theta^{\rm r}  &= \Delta \delta (\vec q, \vec k)  + k_x x - k_y y + \rho (\vec q+\vec k) - \rho (\vec q ), 
\end{align}
where $\Delta \delta (\vec q, \vec k)  = \delta (\vec q+ \vec k) - \delta (\vec q)$. 
In what follows, we will analyze Eq.~\ref{eq:E2} for $|f(\vec q)|$ that is narrowly distributed about $\vec q_0$.

For clarity, we first concentrate on the case of a single frequency~\cite{Miller,Fradkin,Beenakker,Haan,Zhenhua,QDong} valid for plasmon beams with a narrow frequency width.
In this case, $ {\boldsymbol \varepsilon}^{\rm i,r} (\vec q ) \cdot {\boldsymbol \varepsilon}^{\rm i,r} (\vec q + \vec k ) 
= (2 \pi \mathcal{N} q_0 /\omega_0)^2 \cos \varphi$ with $\varphi$ the angle between $\vec q$ and $\vec q + \vec k$.
As a result, $f_E^{\rm i } (\vec q, \vec k) = f_E^{\rm r} (\vec q, \vec k)$ are sharply peaked about $(\vec q = \vec q_0, \vec k =0)$. The center/peak of $F^{\rm i,r} (x, y)$ in the $y$-direction at a given $x$,
denoted $y_c^{\rm i, r} (x) $ 
can be obtained directly via the method of stationary phases: 
\be
[  {\rm d} \theta^{\rm i,r} (\vec q, \vec k, x, y) / {\rm d} k_y  ]_{\vec q=\vec q_0, \vec k =0} = 0   , 
\label{eq:saddle}
\ee
so that for $y = y_c^{\rm i, r}$, the phases $\theta^{\rm i, r}$ in Eq.~\ref{eq:E2} 
add constructively yielding a maximal $F^{\rm i,r} (x, y)$. By directly taking derivatives, Eq.~\ref{eq:saddle} yields the incident and reflected wavepacket center as described in Eq.~\ref{eq:center} of the main text,
with
\begin{align}
\kappa_0 =  [ {\rm d} k_x  / {\rm d} k_y ]_{\vec q = \vec q_0, \vec k = 0}  = - q_{0,y} / q_{0,x} = - \tan \phi  ,
\end{align}
where $k_x = \sqrt{q_0^2 - (q_y + k_y)^2} - q_x$, and $\phi$ denotes the incident angle. For an explicit calculation of $y_c^{\rm i,r} (x) $ using Gaussian wavepackets see~\cite{supp}. This reproduce all our results.

For a discussion of wavepackets possessing a distribution of frequencies, see {\bf SI}~\cite{supp}. 
For narrow frequency distributions, these yield the same results as discussed in the main text.

\onecolumngrid

\newpage

\section*{Supplementary Information for ``Plasmon geometric phase and plasmon Hall shift''}

\twocolumngrid

\subsection{Relation between eigeonmodes $\vec u^+ (\vec q)$ and $\vec u^- (\vec q)$}
The eigenmodes $\vec u^+ (\vec q)$ and $\vec u^- (\vec q)$ in the main text are complex conjugate to each other: $[\vec u^+ (\vec q)]^* = - \vec u^- (-\vec q)$.
This ensures that fields governed by Maxwell's equations are real-valued. To see this, consider a real-valued current field 
\begin{align} 
\vec J (\vec r, t) &  = \frac{1}{2}\big[ \vec u^+ (\vec q) e^{ i \omega_q t - i \vec q \cdot \vec t}- \vec u^- (- \vec q) e^{i (- \omega_q ) t - i (- \vec q) \cdot \vec t}\big]   .
\label{eq:currentfieldwavereal}
\end{align}
The corresponding electric field can be directly obtained from Eq.~\ref{eq:electricfieldwavereal} by applying ${\boldsymbol {\mathcal F}} (\vec q)$,
which is an operator: 
\begin{align} 
\vec E (\vec r, t)  =&~ {\boldsymbol {\mathcal F}} (\vec q) \vec J (\vec r, t)  \nn
=&~ \frac{1}{2} {\boldsymbol {\mathcal F}} ( \vec q) \vec u^+ (\vec q) e^{ i \omega_q t - i \vec q \cdot \vec t}   \nn
& - \frac{1}{2} {\boldsymbol {\mathcal F}} ( \vec q) \vec u^- (- \vec q) e^{i (- \omega_q ) t - i (- \vec q) \cdot \vec t}  \nn
=&~ \frac{1}{2}\big[{\boldsymbol \varepsilon} (\vec q) e^{ i \omega_q t - i \vec q \cdot \vec t}- {\boldsymbol \varepsilon} (- \vec q) e^{i (- \omega_q ) t - i (- \vec q) \cdot \vec t}\big]  \nn
 =&~ {\boldsymbol \varepsilon} (\vec q) \cos ( \omega_q t - \vec q \cdot \vec r ), 
\label{eq:electricfieldwavereal}
\end{align}
where we have noted that ${\boldsymbol \varepsilon} (\vec q) = {\boldsymbol {\mathcal F}} (\vec q) \vec u^{\pm} (\vec q)  $.
As a result, $\vec E (\vec r, t)$ is also a real-valued field since $[\vec u^+ (\vec q)]^* = - \vec u^- (\vec q)$.
In other words, the requirement of real-valued fields guarantees the existence of both $\vec u^+ (\vec q)$ and $\vec u^- (\vec q)$ bands.

\subsection{Envelope for electric field intensity}
In the main text, we have the incident and reflected wavepackets for electric fields:
\begin{align}
{\vec E}^{\rm i,r} (\vec r)  = \int {\rm d} \vec q f (\vec q)  p^{\rm i,r}(\vec q) {\boldsymbol \varepsilon}^{\rm i,r} (\vec q ) 
e^{ \mp i q_x x - i q_y y } ,
\end{align}
and their intensities in real space are squares of their real parts:
\begin{align}
\lb \Re {\vec E}^{\rm i,r} (\vec r) \rb^2  &=  \Re \int  {\rm d} \vec q {\rm d} \vec k~ f_E^{\rm i, r} (\vec q,\vec k) e^{ i \theta^{\rm i,r} (\vec q, \vec k, \vec r) }   \nn
& \times \frac{1}{2}  [ 1 + e^{ i \theta_+^{\rm i,r}(\vec q, \vec r) }  ] ,
\end{align}
where $f_E^{\rm i, r} (\vec q, \vec k) =  | f (\vec q)| |f (\vec q+ \vec k)|  
 {\boldsymbol \varepsilon}^{\rm i,r} (\vec q ) {\boldsymbol \varepsilon}^{\rm i,r} (\vec q + \vec k)$ is a real function of $(\vec q, \vec k)$, and
\begin{align}
\theta^{\rm i}  &=  - k_x x - k_y y + \Delta \delta(\vec q, \vec k), \nn
\theta^{\rm r}  &= + k_x x - k_y y + \Delta \delta(\vec q, \vec k) + \Delta \rho (\vec q, \vec k) , \nn
\theta_+^{\rm i}  &= - 2 q_x x - 2 q_y y + 2 \delta(\vec q), \nn
\theta_+^{\rm r}  &= +2 q_x x - 2 q_y y + 2 \delta(\vec q) + 2 \rho (\vec q) , 
\end{align}
where $\Delta \delta(\vec q, \vec k)  =  \delta (\vec q+ \vec k) - \delta (\vec q)$
and $\Delta \rho(\vec q, \vec k)  =  \rho (\vec q+ \vec k) - \rho (\vec q)$.

Because $f_E^{\rm i, r} (\vec q, \vec k)$ is peaked around $(\vec q = \vec q_0, \vec k =0)$, therefore $\exp [ i \theta_+^{\rm i.r } (\vec q) ] \sim \exp ( \mp 2 i q_x x - 2 i q_y y ) $ is highly oscillatory. This contrasts with the slowly varying envelope function $\exp [ i \theta^{\rm i,r} (\vec q, \vec k, \vec r) ] \sim \exp ( \mp i k_x x - i k_y y )$. As a result, to describe the wave packet centers, we focus on the envelope function to track the slowly varying electric field intensities:
\begin{align}
F^{\rm i,r} (\vec r) =& \int  {\rm d} \vec q {\rm d} \vec k~ f_E^{\rm i, r} (\vec q,\vec k)  \exp [ i \theta^{\rm i,r} (\vec q, \vec k, \vec r) ]  . 
\label{suppeq:envelop}
\end{align}

\subsection{Plasmon Hall shift using Gaussian wavepackets}

The formulation discussed in the main text, reveals a plasmon Hall shift that is insensitive to the details of the profile $f (\vec q)$. For concreteness, in this section we calculate the plasmon Hall shift for a specific wavepacket profile: Gaussian wavepackets.  
To proceed we construct an incident wavepacket composed of $\vec u (\vec q)$ modes, with a fixed frequency $\omega_q = \omega_0$ and a distribution of incident angles $\phi$ which satisfy a Gaussian distribution $f (\phi) = \exp [- (\phi-\phi_0)^2 / 2 \ell^2]$ of width $\ell$.
Here we set $\delta(\phi) = 0$ without loss of generality, since it corresponds to a reference point $y_0 = 0$ (see Eq.~\ref{eq:center} of the main text).
From Eq.~\ref{suppeq:envelop}, the envelope function for incident/reflected electric field intensity is
\begin{align}
F^{\rm i,r} (y)  = & \int {\rm d} \phi {\rm d} \varphi~  f (\phi) f (\phi+\varphi)   p^{\rm i,r} (\phi)^*  p^{\rm i,r} (\phi+ \varphi )    \nn
& \times  {\boldsymbol \varepsilon}^{\rm i,r} (\phi) {\boldsymbol \varepsilon}^{\rm i,r} (\phi+ \varphi) \exp ( - i k_y y  )  ,
\label{suppeq:E2}
\end{align}
where ${\boldsymbol \varepsilon}^{\rm i,r} (\phi) {\boldsymbol \varepsilon}^{\rm i,r} (\phi + \varphi ) =  \varepsilon_0^2 \cos \varphi $
with $\varepsilon_0 =  2 \pi {\cal N}_0  q_0/ \omega_0 $, $k_y = q_0 [\sin (\phi+ \varphi) - \sin \phi]$. Setting 
the incident amplitude $p^{\rm i} (\phi) = 1 $, the 
\begin{align}
p^{\rm r} (\phi) = \frac{i \cos \phi - \eta_0 \sin \phi}{i \cos \phi + \eta_0 \sin \phi}   ,
\end{align} 
with $\eta_0 = \omega_c / \omega_0  $.
In Eq.~\ref{suppeq:E2}, we have assumed a narrow beam profile with $\ell \ll 1$. This allows us to extend the integral limits of $\phi$ and $\varphi$ from $[-\pi, \pi]$ to $(-\infty, \infty)$. In Eq.~\ref{suppeq:E2} we have set $x=0$, since we are interested in the plasmon Hall shift (shift between $y_c^{\rm i}$ and $y_c^{\rm r}$ position) along the boundary $x = 0$. 

Because $(\phi, \varphi)$ contributes mostly  around $(\phi_0, 0)$, we expand $(\phi, \varphi)$ around this point to linear order,
i.e., $k_y \approx q_0 \cos \phi_0 \varphi $ and 
$p^{\rm r} (\phi)^* p^{\rm r} (\phi+ \varphi ) \approx \exp [ 2 i \eta_0 \varphi / (\cos^2 \phi_0 + \eta_0^2 \sin^2 \phi_0)]$.
Using these, the integral in Eq.~\ref{suppeq:E2} can be approximated by the standard Gaussian integral:
\begin{align}
F^{\rm i,r} (y)  \approx  \varepsilon_0^2 \int {\rm d} \phi {\rm d} \varphi~  e^
{- \frac{(\phi-\phi_0)^2+(\phi + \varphi - \phi_0)^2 }{2 \ell^2} - i \varphi Y^{\rm i, r} (y) } \cos \varphi   ,
\label{eq:gaussianF}
\end{align}
where $Y^{\rm i} (y) = q_0 \cos \phi_0 y$ and $ Y^{\rm r} (y) = q_0 \cos \phi_0 y - 2 \eta_0 /( \cos^2 \phi_0 + \eta_0^2 \sin^2 \phi_0)$.
Directly integrating Eq.~\ref{eq:gaussianF}, the envelope functions can be written as

\begin{align}
F^{\rm i,r} (y) \approx &~ \pi \ell ^2 \varepsilon_0^2  
\lp e^{ - \ell^2 [Y^{\rm i,r} (y) - 1]^2 } + e^{- \ell^2 [Y^{\rm i,r} (y) + 1]^2}  \rp \nn
= &~ 2 \pi \ell^2 \varepsilon_0^2  e^{- [
\ell Y^{\rm i,r} (y) ]^2 - \ell^2 + O(\ell^4)}   .
\end{align}
The center for $F^{\rm i} (y)$ and $F^{\rm r} (y)$ are therefore at $Y^{\rm i,r} (y) = 0$, yielding the wavepacket centers at $x=0$ as 
\be
y_c^{\rm i} =  0, \quad  y_c^{\rm r} =  \frac{2 \eta_0}{q_0}\frac{\sec \phi_0}{\cos^2 \phi_0 + \eta_0^2 \sin^2 \phi_0}  ,
\ee
which gives the Hall shift as we discussed in the main text.

\subsection{Magnetoplasmon scattering across a metallic interface}

In the main text, we analyzed reflection at a metal/vacuum boundary. In this section, we will discuss scattering of a plasmon wave at an interface between two uniform regions: 
region 1 and region 2.
The two regions have a sharp interface at $x=0$.
They can have the different $a_0$ and $\omega_c$.
These are modeled by step function: $a_0(x<0) = a_0^{(1)}$, and $a_0(x>0) = a_0^{(2)}$; $\omega_c(x<0) = \omega_{c,1}$, and $\omega_c (x> 0 ) = \omega_{c,2}$. For simplicity, we assume the incident wave is a monochromatic plane wave: ${\boldsymbol {\mathcal E}}_{\rm i} (\vec r, t) = \vec E_{\rm i} (\vec q_{\rm i}) \exp ( i  \omega_{\vec q_{\rm i}} t - i \vec q_{\rm i} \cdot \vec r) $, which excites the mode $ \vec u (\vec q_{\rm i}) \exp ( i \omega_{\vec q_{\rm i}} t - i \vec q_{\rm i} \cdot \vec r)$.
Current conservation at the $x=0$ interface requires
\begin{align}
p_{\rm i} \vec u (\vec q_{\rm i}) e^{ i  \omega_{\vec q_{\rm i}} t }  + 
&  p_{\rm r} \vec u (\vec q_{\rm r}) e^{i \omega_{\vec q_{\rm r}} t }  
= p_{\rm t} \vec u (\vec q_{\rm t})  e^{ i \omega_{\vec q_{\rm t}} t }. 
\label{suppeq:currentcontinue}
\end{align}
Here $\vec q_{\rm i} = (q_{1,x}, q_y)$, $\vec q_{\rm r} = (- q_{1,x}, q_y)$ and $\vec q_{\rm t} = (q_{2,x}, q_y)$. $q_y$ is conserved because of translational symmetry in the $y$-direction. Applying energy conservation $a_0^{(1)} |\vec q_{\rm r} | + \omega_{c,1}^2 = a_0^{(2)} |\vec  q_{\rm t}| + \omega_{c,2}^2$ yields $q_{1,x} \rightarrow - q_{1,x}$ for the reflected wave, and $ a_0^{(2)} (q_{2,x}^2+q_y^2)^{1/2} + \omega_{c,2}^2 = a_0^{(1)} (q_{1,x}^2+q_y^2)^{1/2} + \omega_{c,1} ^2$ for the transmitted wave.

Substituting the eigenmode
\be
 \vec u_{1,2} (\vec q)
= \frac{{\cal N}_{1,2}}{q}
\lp \begin{array}{c}
  - i q_x + \eta_{1,2} q_y \\
  - i q_y - \eta_{1,2} q_x
\end{array} \rp   ,
\ee
for the two regions respectively, as well as the incident angle $\phi_1$, reflection angle $\pi - \phi_1$, transmission angle $\phi_2$, and the corresponding $\eta_{1,2}$ into the boundary condition Eq.~\ref{suppeq:currentcontinue},
we have
\begin{align}
\frac{ p_{\rm r }}{ p_{\rm i} } & =   
\frac{- i (\eta_1 - \eta_2) \cos(\phi_1 - \phi_2) + (1 - \eta_1 \eta_2) \sin (\phi_1 - \phi_2) }
{- i (\eta_1 - \eta_2) \cos (\phi_1 + \phi_2) + (\eta_1 \eta_2 - 1 )\sin (\phi_1 + \phi_2) }  ,  \nn
\frac{ {\cal N}_2 p_{\rm t} }{ {\cal N}_1 p_{\rm i} } & =  
\frac{ (\eta_1^2 - 1 ) \sin 2\phi_1 }
{- i (\eta_1 - \eta_2) \cos (\phi_1 + \phi_2) + (\eta_1 \eta_2 - 1 )\sin (\phi_1 + \phi_2) }  ,
\label{eq:newf}
\end{align}
where ${\cal N}_1$ and ${\cal N}_2$ are normalization constants in region 1 and 2, respectively.
When $\eta_1 = \eta_2$, this gives the Fresnel equations:
\be
\frac{ p_{\rm r} }{ p_{\rm i} }  =   
- \frac{  \sin (\phi_1 - \phi_2)  }
{ \sin (\phi_1 + \phi_2) }  ,  \quad
\frac{ p_{\rm t} }{ p_{\rm i} }  =  
\frac{  \sin 2\phi_1 }
{\sin (\phi_1 + \phi_2)  }  .
\ee
When $\eta_1 \neq \eta_2$, there are deviations from the conventional Fresnel equations (see Eq.~\ref{eq:newf}). This is because the current density textures in region 1 and region 2 are not perfectly matched.

\subsection{Wavepacket center for current and power density}

In the main text, we illustrated the plasmon Hall shift by considering wavepackets of electric field intensity. As we now discuss, wavepackets of other quantities can also experience plasmon Hall shifts. For example, the current density of the wavepacket along its propagation direction can be written as
\begin{align}
J^{\rm i, r}  (\vec r) =&~ \Re 
\int {\rm d} \vec q ~ f (\vec q)  \vec u^{\rm i, r} (\vec q ) \cdot (\vec q^{\rm i, r} / q) \exp [i \rho^{\rm i,r} (\vec q ) ] \nn
& \times  \exp ( \mp i q_x x - i q_y y  )  \nn
=&~ \Re  \int {\rm d} \vec q ~ {\cal N} f (\vec q)  \exp [i \rho^{\rm i,r} (\vec q ) ] \nn
& \times  \exp ( \mp i q_x x - i q_y y  )  ,
\label{suppeq:currentdensity}
\end{align}
where $(\vec q^{\rm i, r} / q)$ is the unit vector along the propagating direction of mode 
$\vec u^{\rm i, r} (\vec q )\exp ( \mp i q_x x - i q_y y )$.
We find that the center of the current density (wavepacket) is the same with $y_c^{\rm i, r} (x)$ determined in Eq.~\ref{eq:center} of the main text. To see this, note that the center of the wavepacket can be determined by the method of stationary phases (see main text, and ``Methods'') directly on Eq.~\ref{suppeq:currentdensity} as 
\be
\left. {\rm d} [ \mp q_x x - q_y y_c^{\rm i,r} + \rho^{\rm i,r} (\vec q) ] / {\rm d} q_y  \right|_{\vec q=\vec q_0}  = 0. 
\label{eq:currentdensitysaddle}
\ee
Solving Eq.~\ref{eq:currentdensitysaddle} reduces to the expression in Eq.~\ref{eq:center} of the main text, allowing the current density (along propagation direction) to also experience the same plasmon Hall shift (Eq.~\ref{eq:deltay}) as discussed in the main text. 

In the same fashion, we write the wavepacket for the power density  ${\cal I}^{\rm i, r}(\vec r) = \vec E^{\rm i,r} (\vec r) \cdot \vec J^{\rm i,r} (\vec r)$ as
\begin{align}
{\cal I}^{\rm i, r} (\vec r)  =
&~ \Re  \int  {\rm d} \vec q {\rm d} \vec k~ f_{\cal I} (\vec q, \vec k) \exp [ i \theta^{\rm i,r} (\vec q, \vec k, \vec r) ]  ,
\label{suppeq:power}
\end{align}
where
$f_{\cal I} (\vec q, \vec k) = 2\pi {\cal N}^2 q_0^2 |f (\vec q)| |f (\vec q+ \vec k)| \cos \varphi /\omega_q $ with $\varphi$ the angle between $\vec q$ and $\vec q + \vec k$ for a monochromatic frequency. 
This gives a sharp peak for $f_{\cal I} (\vec q, \vec k)$ at
$(\vec q = \vec q_0, \vec k = 0)$.
The center of ${\cal I}^{\rm i, r} (\vec r)$ in real space is again the same saddle point $y_c^{\rm i, r} (x)$ determined by Eq.~\ref{eq:saddle} of the main text. This means that the power density also experiences the same plasmon Hall shift as discussed in Eq.~\ref{eq:deltay} of the main text.

\subsection{Stationary phase for multiple frequencies}

In the main text, we concentrated on plasmon beams with a single frequency. This can be achieved by exciting plasmons via a laser beam with narrow line width; these sources are to all intents and purposes practically monochromatic. Nevertheless, we note parenthetically that the analysis in ``Methods'' can be extended to cases when the frequency has a distribution with a single peak.
We first consider cases in which $\omega_q$ can only take discrete values:
$\omega_q \in \{  \omega_n |  n \in \mathbb{N} \}$,
with $\omega_0 = \omega (q_0)$ having the largest amplitude. Focussing on the maximum amplitude $\omega_0$ distribution, we can follow the same analysis as in ``Methods'' 
to track the trajectory of the plasmon beam as Eq.~\ref{eq:center} of the main text.

For a continuous distribution $f (\vec q)$ over multiple frequencies, a more elaborate analysis taking into account the distribution's shape and principal directions~\cite{Gray} can be employed. Finite width (in frequency space) corrections to quantities such as $\kappa_0$ occur only when the frequency distribution is correlated with its angle distribution $\phi$, and are typically small.

\end{document}